\documentclass[12pt]{article}

\usepackage{float}
\usepackage[round,comma,authoryear]{natbib}
\usepackage{amsmath, amssymb, amsfonts}
\RequirePackage{amsthm}
\usepackage{algorithm}
\usepackage{algorithmic}
\usepackage{hyperref}
\usepackage{amsmath, amssymb, amsfonts} 
\usepackage{graphicx, subfigure} 
\usepackage{booktabs}            
\usepackage{multirow}            
\usepackage{color,soul}
\graphicspath{{./figs/}}

\renewcommand\appendix{\par
  \setcounter{section}{0}
  \setcounter{subsection}{0}
  \setcounter{figure}{0}
  \setcounter{table}{0}
  \renewcommand\thesection{Appendix \Alph{section}}
  \renewcommand\thefigure{\Alph{section}\arabic{figure}}
  \renewcommand\thetable{\Alph{section}\arabic{table}}
}

\title{Detecting Epistatic Selection with Partially Observed Genotype Data Using Copula Graphical Models }

\author{P. Behrouzi\\
  Wageningen University and Research Centre\\
  \texttt{pariya.behrouzi@wur.nl}
  \and E. C. Wit\\
  University of Groningen\\
  \texttt{e.c.wit@rug.nl}
  }

\begin{document}

\maketitle

\begin{abstract}
  Recombinant Inbred Lines derived from divergent parental lines can display extensive segregation distortion and long-range linkage disequilibrium (LD) between distant loci. These genomic signatures are consistent with epistatic selection during inbreeding. Epistatic interactions affect growth and fertility traits or even cause complete lethality. Detecting epistasis is challenging as multiple testing approaches are under-powered and true long-range LD is difficult to distinguish from drift. 
  Here we develop a method for reconstructing an underlying network of genomic signatures of high-dimensional epistatic selection from multi-locus genotype data. The network captures the conditionally dependent short- and long-range LD structure and thus reveals {\textquotedblleft aberrant\textquotedblright} marker-marker associations that are due to epistatic selection rather than gametic linkage. The network estimation relies on penalized Gaussian copula graphical models, which accounts for a large number of markers $p$ and a small number of individuals $n$.
  A multi-core implementation of our algorithm makes it feasible to estimate the graph in high-dimensions also in the presence of significant portions of missing data. We demonstrate the efficiency of the proposed method on simulated datasets as well as on genotyping data in \emph{A.thaliana} and maize. In addition, we implemented the method in the R package {\tt netgwas} which is freely available at \url{https://CRAN.R-project.org/package=netgwas}.

 \end{abstract}

\section{Introduction}
The Recombinant Inbred Lines (RILs) study design is a popular tool for studying the genetic and environmental basis of complex traits. It has become a valuable resource in biomedical and agricultural research. Many panels of RILs exist in a variety of plant and animal species. RILs are typically derived from two divergent inbred parental strains, but multi-parental RILs have been recently established in A. thaliana, Drosophila, and mouse originating from four or eight inbred parents \citep{broman2005genomes, gibson2002enabling, threadgill2002genetic}. The construction of RILs is not always straightforward: low fertility, or even complete lethality, of lines during inbreeding is common, particularly in natural outcrossing species \citep{rongling1999multiplicative, wu2000quantitative}, and can severely bias genotype frequencies in advanced inbreeding generations. These genomic signatures are indicative of epistatic selection having acted on entire networks of interacting loci during inbreeding, with some combinations of parental alleles being strongly favored over others. 

Recently, \citet{colome2015signatures} studied two-loci epistatic selection in RILs. However, the reconstruction of multi-loci epistatic selection network has received little attention by experimentalists. One important reason is that large numbers of potentially interacting loci are methodologically and computationally difficult. One intuitive approach to this problem is to perform an exhaustive genome scan for pairs of loci that show significant long-range LD or pair-wise segregation distortion, and then try to build up larger networks from overlapping pairs. \citet{torjek2006segregation}, for instance, employed this idea for the detection of possible epistasis by testing for pairwise segregation distortion. The drawback of such an approach is that hypothesis testing in the genome-scale is heavily underpowered, so that weak long-range LD will go undetected, especially after adjusting for multiple testing. Furthermore, pair-wise tests are not, statistically speaking, consistent \citep{whittaker2009graphical} when two conditionally independent loci are mutually dependent on other loci, and may, therefore, lead to incorrect signatures. 

In order to overcome some of these issues, we shall argue that the detection of epistatic selection in RIL genomes can be achieved by inferring a high-dimensional graph of conditional dependency relationships among loci. Technically, this requires estimating a sparse adjacency matrix from a large number of discrete ordinal marker genotypes, where the number of markers $p$ can far exceed the number of individuals $n$. The estimated conditional independence graph captures the conditionally dependent short- and long-range LD structure of RIL genomes, and thus provides a basis for identifying associations between distant markers that are due to epistatic selection rather than gametic linkage. 

In this paper, we introduce an efficient method to perform this estimation. To this end, we propose an $\ell_1$ regularized latent graphical model, which involves determining the joint probability distribution of discrete ordinal variables. The genotype data contain information on measured markers in the genome which are generally coded as the number of paternal or maternal alleles, for instance $0, 1$ and $2$ for a heterozygous population in a diploid species. Sklar's theorem shows that any $p$-dimensional joint distribution can be decomposed into its $p$ marginal distributions and a copula, which describes the dependence structure between $p$-dimensional multivariate random variable \citep{nelsen1999introduction}. Various statistical network modeling approaches have been proposed for inferring high-dimensional associations among non-Gaussian variables \citep{liu2009nonparanormal, liu2012high, dobra2011copula, mohammadi2016bayesian}. The above-mentioned models have some limitations; the first two methods cannot deal with missing data, and the last two are computationally expensive since their inference is based on a Bayesian approach. Studying the conditional relationships between ordinal discrete variables is complicated since we are faced with two challenges. First, general dependence structure can be very complicated, way beyond the pairwise dependencies of a normal variate. Second, univariate marginal distributions cannot be adequately described by simple parametric models. To handle the first challenge we used a Gaussian copula; effectively transforming each of the marginal distributions to a standard Gaussian distribution. To address the second challenge, we treat the marginal distributions as nuisance parameters that we estimate non-parametrically.
 
This paper is organized as follows. In section \ref{epistasis}, we describe the genetic background on  epistatic selection. Section \ref{GMEpi} explains the model and introduces the Gaussian copula graphical model connecting the observed marker data with the underlying latent genotype. In addition, we explain how to infer the conditional dependence relationships between multi-loci in genome-wide association studies (GWAS), using the $\ell_1$ regularized Gaussian copula framework. In section \ref{simulation}, we investigate the performance of the proposed method in terms of precision matrix estimation. Also, we compare the performance of our proposed method with alternative approaches in terms of graph recovery. We have implemented the method in the R package {\tt netgwas} \citep{behrouzi2017netgwas}. 
In Section $5$, we aim to reveal genomic regions undergoing selection in two species. We apply our proposed method to the well-studied cross Col $\times$ Cvi in \emph{Arabidopsis thaliana} in section \ref{arabi}, and to high-dimensional $B73 \times Ki11$ genotype data from Maize Nested Association Mapping (NAM) populations in section \ref{maize}, where $1106$ genetic markers were genotyped for $193$ individuals. 

\section{Genetic background of epistatic selection}
\label{epistasis}
Two alleles at locations $l_1$ and $l_2$ are said to act additively if the effect of the first allele on the phenotype does not depend on the state of the second allele, and vice versa. On the other hand, epistasis refers to the interaction of alleles at different loci on that phenotype. Epistasis occurs when the joint effect of a particular pair of loci is different from what would be expected under additivity. In this section, we provide the genetic background on epistatic selection, i.e. the case in which the phenotype of interest is survival.

\subsection{Meiosis}
Sexual reproduction involves meiosis. Meiosis is a form of cell division that produces gametes (egg/ sperm). During this process, the arms of homologous chromosomes can recombine, which involves the sequential alignment of genetic material from the maternal and paternal chromosomes. As a result, offspring can have different combinations of alleles than their parents. Genetic markers, regions of DNA, that physically located close together on the same chromosome have a tendency to be transmitted together in meiosis. This tendency is called linkage. Loci on different chromosomes have no linkage and they assort independently during meiosis. Statistically speaking, genetic linkage means observing dependence between markers that are physically close together on the same chromosome. 

Linkage disequilibrium refers to the co-inheritance of alleles at different but functionally related loci. If two loci are in linkage equilibrium, it means that they are inherited completely independently in each generation. If two loci are in linkage disequilibrium, it means that certain alleles of each loci are inherited together more or less often than would be expected by chance. This may be due to actual genetic linkage when the loci are located on the same chromosome. However, if loci are located on different chromosomes, this is due to some form of functional interaction where certain combinations of alleles at two loci affect the viability of potential offspring. 
\begin{figure}[t]
\centering
\includegraphics[width=0.27\textwidth]{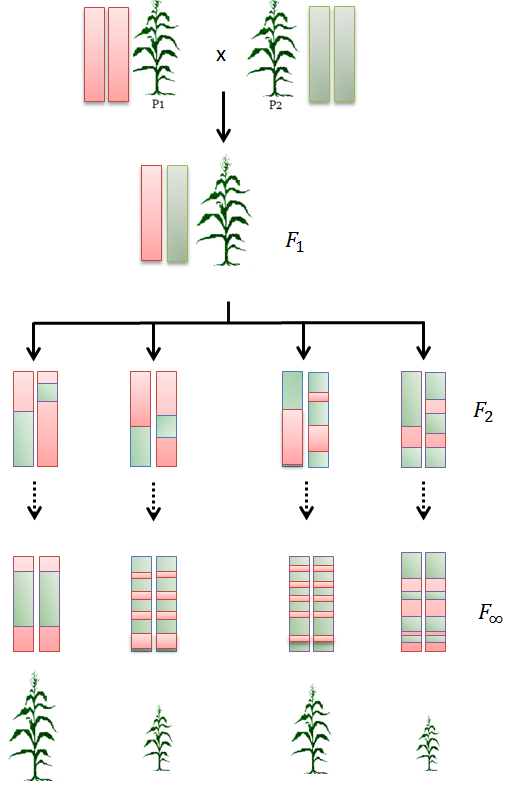}
\caption{\scriptsize The production of recombinant inbred lines (RILs) by repeated selfing. \label{RIL}}
\end{figure}
\subsection{Recombinant Inbred Lines}
Recombinant inbred lines (RILs) are typically derived by crossing two inbred lines followed by repeated generations of selfing or sibling mating to produce an inbred line whose genome is a mosaic of its parental lines. For instance, if a diploid allele of parent $P1$ is labeled $A$ and that of $P2$ is labeled $B$, then from generation to generation these alleles recombine and produce different genotypes. For example, due to inbreeding, $P1$ has a homozygous genotype, say $A/A$ (red in Figure \ref{RIL}), at each locus, while $P2$ has homozygous genotype, say $B/B$ (green in Figure \ref{RIL}), at each locus. Crossing $P1$ and $P2$ produces an $F1$ generation with a $A/B$ genotype at each locus. The subsequent $F2$ followed by repeated selfing results in a genome in the obtained offspring that is a mosaic of the two parental allele combinations (see Figure \ref{RIL}).

\subsection{Genome-wide association study}
A pure RIL would result in one of two genotype at each locus: either A/A or B/B. However, in practice in a two-way RIL (see Figure \ref{RIL}), the genotype state of an offspring at a given loci comes either from parent $1$, parent $2$, or in a small fraction of cases from both parental alleles. For instance, in a diploid organism the genotype states at each chromosomal position are either $0$ (homozygous $AA$ from one parent), $2$ (homozygous $BB$ for the other parent), or $1$ which defines the heterozygous genotype $AB$. The routine way of coding a diploid genotype data is to use $\{0, 1, 2\}$ to represent $\{AA, AB, BB\}$, respectively, where we do not distinguish AB and BA. 

A complete genome consist of billions of loci, many of which do not vary between individuals in a population. Clearly those loci are inherited without change from generation to generation, unless some mutation occurs. Single nucleotide polymorphisms (SNPs) are loci where the genotype does vary, either homozygously $\{0, 2\}$ or heterozygously $\{0, 1, 2\}$, considering diploid organisms. Genome-wide association studies measure thousands of SNPs along the genome, resulting for each individual in a partially ordered vector $Y=(Y_1,\ldots,Y_p)$ of $p$ markers on the genotype: within each chromosome the markers are ordered, but between chromosomes there is no natural ordering. The component $Y_j$ for an individual indicates the ancestral genotype value for marker $j$, i.e., either $0$ or $2$ for homozygous populations and $0, 1$ or $2$ for heterozygous diploid populations.  

Genome-wide association studies (GWAS) were designed to identify genetic variations that are associated with a complex trait. In a GWAS, typically a small number of individuals are genotyped for hundreds of thousands of SNPs. SNP markers are naturally ordered along the genome with respect to their physical positions. Nearby loci can be highly correlated due to genetic linkage. Moreover, linkage groups typically correspond to chromosomes. 

\begin{figure}[t]
\centering
\includegraphics[width=0.6\textwidth]{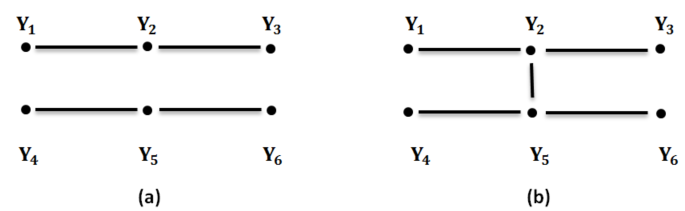}
\caption{\scriptsize Cartoon representation of $6$ markers on $2$ different chromosomes where $Y_1$, $Y_2$ and $Y_3$ belong to chromosome $1$ and $Y_4$, $Y_5$ and $Y_6$ belong to chromosome $2$. Conditional independence relationships between markers (a) in the absence of epistatic selection, in other words markers on different chromosomes segregate independently, and (b) in the presence of epistatic selection. Markers $2$ and $5$ have an epistatic interaction, resulting in long-range linkage disequilibrium. \label{simpleExpEpis} }
\end{figure}
\subsection{Epistatic phenotype}
Epistasis is typically defined with respect to some explicit phenotype, such as the shape of the comb in a chicken or the flower color in peas \citep{Batesonw}. In RILs the phenotype we consider, however, is not explicit, but implicit: the viability of the particular genetic recombination of the parental lines results in the presence or absence of such recombination in the progeny.

In the construction of RILs from two divergent parents certain combinations of genotypes may not function well when brought together in the genome of the progeny, thus resulting in sterility, low fertility, or even complete lethality of lines during inbreeding. This can result in recombination distortion within chromosomes, short-range linkage disequilibrium, or segregation distortion across chromosomes, also called long-range linkage disequilibrium (lr-LD). Thus, the genomic signatures of epistatic selection will appear as interacting loci during inbreeding, whereby some combinations of parental alleles will be strongly favored over others.  

It has long been recognized that detecting the genomic signatures of such high-di\-men\-sio\-nal epistatic selection can be complex, involving multiple loci \citep{wu2000quantitative, Mather}. The detection of high-dimensional epistatic selection is an important goal in population genetics. The aim here is to propose a model for detecting genomic signatures of high-dimensional epistasis selection during inbreeding. 


\section{Graphical model for epistatic selection}
\label{GMEpi}
If meiosis is a sequential markov process, then in the absence of epistatic selection the genotype $Y$ can be represented as a graphical model \citep{lauritzen1996graphical} for which the conditional independence graph corresponds with a linear representation of the chromosome structure (see Figure \ref{simpleExpEpis}a). However, in the presence of epistatic selection, the conditional independence of non-neighboring markers may become undone. This could result, for example, in an underlying conditional independence graph as shown in Fig \ref{simpleExpEpis}b, which shows $6$ markers on $2$ chromosomes whereby markers $2$ and $5$ have an epistatic interaction that affects the viability of the offspring. 

In the next section, we define a convenient semi-parametric model, which can easily be generalized to large sets of markers. We assume a known genetic map, and let $y_j^{(i)}$ $j=1,\ldots,p$; $i=1, \ldots, n$ denote the genotype of $i$\emph{th} individual for $j$\emph{th} SNP marker. The observations $y_j^{(i)}$ arise from $\{0, 1, \ldots, k_j-1 \}, k_j \geq 2$ discrete ordinal values. In the genetic set-up, $k_j$ is the number of possible distinct genotype states at locus $j$. For instance, in a tetraploid species $k_j$ takes either the value $2$ in a homozygous population, or $5$ in a heterozygous population. 

\begin{figure}[t]
\centering
\includegraphics[width=0.84\textwidth]{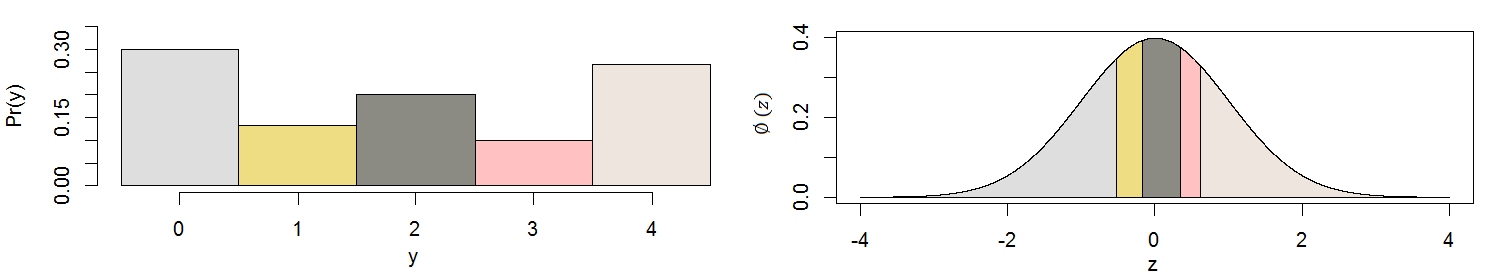}
\caption{ \scriptsize
Relation between j\emph{th} true latent values, $z_j$, and the j\emph{th} observed variable, $y_j$. Here, $k=5$ corresponding with the distinct genotype states in tetraploid species, which contain four copies of the same chromosome.
\label{marginals}}
\end{figure}

\subsection{Gaussian copula graphical model}
\label{GCGM}
A copula is a multivariate cumulative distribution function with uniform mar\-gi\-nals. Sklar's theorem shows that any $p$-dimensional joint distribution can be decomposed into its $p$ marginal distributions, $F_j$, and a copula. This decomposition suggests that the copula captures the dependence structure between $p$ multivariate data. Let $y$ be the collection of all $p$ measured genetic markers across a genome. A genetic marker $Y_j$ takes a finite number of ordinal values from $\{0, 1, \ldots, k_j-1\}$, with $k_j \ge 2$. The marker $Y_j$ is defined as the discretized version of a continuous variable $Z_j$, which cannot be observed directly. The variable $Z$ helps us to construct the joint distribution of $Y$ as follows:
\[
Z \sim N_p(0, \Theta ^{-1}), 
\]  
and the Gaussian copula modeling can be expressed as 
\[
Y_j= F^{-1}_j(\Phi(Z_j)), 
\]
where $\Theta^{-1}$ is a correlation matrix for the Gaussian copula, and $F_j$ denotes the univariate distribution of $Y_j$. We write the joint distribution of $Y$ as
\[
P(Y_1 \leq y_1, \ldots, Y_p \leq y_p) = C(F_1(y_1), \ldots, F_p(y_p) | \Theta ),
\]
where 
\begin{equation}
C(F_1(y_1), \ldots, F_p(y_p )| \Theta^{-1} )= \Phi_{\Theta^{-1}} ( \Phi^{-1}(F_1(y_1)),\ldots, \Phi^{-1}(F_p(y_p))).
\label{jointDistribution}
\end{equation}
Here, $\Phi$ defines the CDF of the standard normal distribution and $\Phi_{\Sigma}$ is the CDF of $N_p(0, \Sigma)$. 

Our aim is to reconstruct the underlying conditional independence graph by using the continuous latent variable $Z$. Typically the relationship between the $j$th marker $Y_j$ and the corresponding $Z_j$ is expressed through a set of cut-points $-\infty= c_{j, 0} < c_{j, 1} < \ldots < c_{j, k_j-1}  < c_{j, k_j} = \infty$, where $c_{j, y+1}= \Phi^{-1}(F_j(y))$. Thus, $y_j^{(i)} $ can be written as follows: 
\begin{eqnarray}
\label{dat}
y_j^{(i)} = \sum_{l=0}^{k_j-1} l \times I_{ \{ c_{j, l} < z_j^{(i)} \le c_{j, l+1} \}}, 	\qquad i= 1, 2,\ldots, n.
\end{eqnarray} 
The j-\emph{th} observed variable $y_j^{(i)}$ takes its value according to latent variable $z_j^{(i)}$. Figure \ref{marginals} displays how the observed data can be obtained from the latent variable using the Gaussian copula. 

Assuming $\mathcal{D}_F(y)= \{z_j^{(i)} \in \mathbb{R} | c_{j, y_j^{(i)}} < z_j^{(i)} \le c_{j,y_j^{(i)}+1} \}$, the likelihood function of a given graph with a precision matrix $\Theta$ and marginal distributions $F$ is defined as 
\begin{equation}
\label{joint likelihood}
L_y(\Theta, F) = \int_{\mathcal{D}_F(y)} p(z \ | \ \mathbf{\Theta}) dz.
\end{equation}

\subsection{$\ell_1$ penalized inference of Gaussian copula graphical model}
\label{inference}

Let $y^{(1)}, \ldots, y^{(n)}$ be i.i.d sample values from the above Gaussian copula distribution. Copulas allow one to learn the marginals $F_j$ separately from the dependence structure of $p$-variate random variables. In the proposed copula modeling, we estimate the correlation matrix--the parameter of interest--with a Gaussian copula, and treat the marginals as nuisance parameters and estimate them non-parametrically through the empirical distribution function $\hat{F}_j(y)= \frac{1}{n} \sum_{i=1}^{n} I\{y_j^{(i)} \leq y\}$. Hence, in the likelihood (\ref{joint likelihood}) the precision matrix of the Gaussian copula, $\Theta$, is the only parameter to estimate, as we replace $\mathcal{D}_F(y)$ by $\mathcal{D}_{\widehat{F}}(y)$ which we will simply indicate as $\widehat{\mathcal{D}}$.

We impose a sparsity penalty on the elements of the precision matrix $\mathbf{\Theta}$ using an $\ell_1$-norm penalty \citep{abegaz2015copula, friedman2008sparse}. Genetically speaking, this sparsity is sensible as we expect a priori only a small number of pairs of LD markers beyond the neighbouring markers. The $\ell_1$ penalized log-likelihood function of genetic markers can be written as
\begin{equation}
\mathcal{\ell}_\mathbf{y}^p ( \mathbf{\Theta}) \approx \frac{n}{2} \log \ | \  \mathbf{\Theta} \ | \ - \frac{1}{2} \sum_{i=1}^{n} \int_{\mathcal{\widehat{D}} } \ldots \int z^{(i)t} \mathbf{\Theta} z^{(i)} dz^{(i)}_1 \ldots dz^{(i)}_p  - \lambda ||\Theta||_1,			   	
\label{likelihood}  
\end{equation}
where $z^{(i)} = (z_1^{(i)}, \ldots, z_p^{(i)})^t$. The maximum $\widehat{\Theta}_\lambda$ of  this log-likelihood function has no closed form expression. To address this problem we introduce a penalized EM-algorithm.

The penalized EM algorithm proceeds by iteratively computing in the E-step the conditional expectation of joint log-likelihood, and optimizing this conditional expectation in the M-step. Assuming that ${\mathbf{\widehat{\Theta}}_\lambda ^{(m)}}$ is the updated approximation of $ \widehat{\mathbf{\Theta}}_\lambda$ in the M-step, then in the E-step the conditional expectation of the joint penalized log-likelihood given the data and $\widehat{ \mathbf{\Theta}}^{(m)}$ is determined.

\begin{align}	
\label{E-step}
Q(\mathbf{ \mathbf{\Theta}} \  | \ \widehat{\mathbf{ \mathbf{\Theta}}}^{(m)}) = & E_Z[\sum_{i=1}^{n} \log p(Z^{(i)}  |   \mathbf{ \mathbf{\Theta}} ) | y^{(i)}, \widehat{ \mathbf{ \mathbf{\Theta}}}^{(m)}, \widehat{\mathcal{D}} ] \nonumber \\
	  				 = & \frac{n}{2}[\log  |   \mathbf{ \mathbf{\Theta}}  |  -tr(\frac{1}{n} \sum_{i=1}^{n} E_{Z^{(i)}}(Z^{(i)} Z^{(i)t}  |  y^{(i)}, \widehat{ \mathbf{ \mathbf{\Theta}}}^{(m)}, \widehat{\mathcal{D}} )  \mathbf{ \mathbf{\Theta}}) -p \log(2\pi)]	 ,  
\end{align}
and 
\begin{equation}
Q_\lambda (\mathbf{ \mathbf{\Theta}} \  | \ \widehat{\mathbf{ \mathbf{\Theta}}}^{(m)}) = Q(\mathbf{ \mathbf{\Theta}} \  | \ \widehat{\mathbf{ \mathbf{\Theta}}}^{(m)}) - \lambda || \Theta || _1 .
\nonumber
\end{equation}
In this equation we still need to evaluate $\bar{R}= \frac{1}{n} \sum_{i=1}^{n} E(Z^{(i)} Z^{(i)t} \ | \ y^{(i)}, \widehat{ \mathbf{\Theta}}^{(m)}, \widehat{\mathcal{D}}) $, 
which we do via one of the two following approaches.

\setlength{\parindent}{0ex} \textbf{A. Monte Carlo Gibbs sampling of latent covariance.} In the Gibbs sampling technique we randomly generate for each sample $Y^{(i)}$ a number of Gibbs samples $Z_\star^{(i)1}, \ldots, Z_\star^{(i)N} $ from a p-variate truncated normal distribution, whose boundaries come from the cut-points of $Y^{(i)}$, as implemented in the R package {\tt tmvnorm} \citep{geweke2005contemporary}. Let 
 \[
   Z_\star^{(i)}=
  \left[ {\begin{array}{c}
   Z_\star^{(i)1}\\   \vdots    \\ Z_\star^{(i)N} \      \end{array} } \right] \in \mathbb{R}^{N \times p},
\]
represent the Gibbs samples for each sample in the data. The expected individual covariance matrix $R_i = E(Z^{(i)} Z^{(i)t} | y^{(i)}, \widehat{\Theta}^{(m)},\widehat{\mathcal{D}})$ can then be estimated as
\[
  \widehat{R}_i = \frac{1}{N} Z_\star^{(i)t}  Z_\star^{(i)}. 
\]
To estimate $\bar{R}$ we take the average of the individual expectation $\widehat{\bar{R}} = \frac{1}{n} \sum\limits_{i=1}^{n} \widehat{R}_i$. We remark that $\widehat{\bar{R}}$ is a positive definite matrix with probability one as long as $N \ge \frac{p}{n}$, since for the $l$th row of $Z_\star^{(i)}$, we have that $Z_{\star l}^{(i)t} Z_{\star l}^{(i)}$ is a rank one, non-negative definite matrix with probability one and, therefore, $\widehat{\bar{R}}$ is of full rank and strictly positive definite, with probability one. In practice, we noticed that the Gibbs sampler needs only few burn-in samples, and $N = 1000$ sweeps is sufficient to calculate the mean of the conditional expectation accurately [more details  in the supplementary material].

\setlength{\parindent}{0ex} \textbf{B. Approximation of the conditional expectation.} Alternatively, we use an efficient approximate estimation algorithm \citep{guo2015graphical}. The variance elements in the conditional expectation matrix can be calculated through the second moment of the conditional $z_{j}^{(i)} \ |\ y^{(i)}$, and the rest of the elements in this matrix can be approximated through $E(Z_{j}^{(i)} Z_{j'}^{(i)} \ | \ y^{(i)}; \widehat{\Theta}, \widehat{\mathcal{D}} ) \approx E(Z_{j}^{(i)}\ | \ y^{(i)}; \widehat{\Theta}, \widehat{\mathcal{D}})$ $E(Z_{j'}^{(i)} \ | \ y^{(i)}; \widehat{\Theta}, \widehat{\mathcal{D}})$ using mean field theory \citep{peterson1987mean}. The first and second moment of $z_{j}^{(i)} | y^{(i)}$ can be written as 
\begin{equation}
\label{firstMoment}
E(Z_{j}^{(i)} \ | \ y^{(i)}, \widehat{\Theta}, \widehat{\mathcal{D}}) = E[E(Z_{j}^{(i)} \ |\ z_{-j}^{(i)}, y_{j}^{(i)}, \widehat{\Theta}, \widehat{\mathcal{D}}) \ | \ y^{(i)}, \widehat{\Theta}, \widehat{\mathcal{D}}],
\end{equation}
\begin{equation}
\label{secondMoment}
E((Z^{(i)}_{j})^2 \ | \ y^{(i)}, \widehat{\Theta}, \widehat{\mathcal{D}}) = E[E((Z^{(i)}_{j})^2 \ |\ z_{-j}^{(i)}, y_{j}^{(i)}, \widehat{\Theta}, \widehat{\mathcal{D}}) \ | \ y^{(i)}, \widehat{\Theta}, \widehat{\mathcal{D}}],
\end{equation}
where $z^{(i)}_{-j} = (z^{(i)}_{1},\ldots, z^{(i)}_{j-1}, z^{(i)}_{j+1}, \ldots, z^{(i)}_{p} )$. The inner expectations in (\ref{firstMoment}) and (\ref{secondMoment}) are relatively straightforward to calculate. $z_{j}^{(i)} \ | \ z_{-j}^{(i)}, y_j^{(i)}$ follows a truncated Gaussian distribution on the interval $[c^{(j)}_{y^{(i)}_{{j}}} , c^{(j)}_{y^{(i)}_{j} + 1}]$ with parameters $\mu_{i,j}$ and $\sigma_{i,j} ^2$ given by
\[
\mu_{ij} = \widehat{\mathbf{\Sigma}}_{j,-j} \widehat{\mathbf{\Sigma}}^{-1}_{-j,-j} z^{(i)t}_{-j},
\]
\[
\sigma_{i,j} ^2 = 1- \widehat{\mathbf{\Sigma}}_{j,-j} \widehat{\mathbf{\Sigma}}^{-1}_{-j,-j} \widehat{\mathbf{\Sigma}}_{-j,-j}.
\]
Let $r_{k, l}= \frac{1}{n} \sum_{i=1}^{n} E(Z^{(i)}_k Z^{(i)}_l \ | \ y^{(i)}, \widehat{ \mathbf{\Theta}}, \widehat{\mathcal{D}} )$  be the $(k,l)$-th element of empirical correlation matrix $\bar{R}$, then to obtain the $\bar{R}$ two simplifications are required.
 
\begin{align}
E(Z^{(i)}_k Z^{(i)t}_l \ | \ y^{(i)},\widehat{ \mathbf{\Theta}}, \widehat{\mathcal{D}} ) & \approx 
	E(Z^{(i)}_k \ | \ y^{(i)},\widehat{ \mathbf{\Theta}} , \widehat{\mathcal{D}}) E(Z^{(i)}_l \ | \ y^{(i)},\widehat{ \mathbf{\Theta}} , \widehat{\mathcal{D}} ) \quad \mbox{if $1 \leq k$ $\neq$ $l$ $\leq p$},  \nonumber \\
E(Z^{(i)}_k Z^{(i)t}_l \ | \ y^{(i)},\widehat{ \mathbf{\Theta}}, \widehat{\mathcal{D}})  & = 	E((Z^{(i)}_k)^2 \ | \ y^{(i)}, \widehat{ \mathbf{\Theta}}, \widehat{\mathcal{D}}) \qquad \qquad \qquad \qquad \mbox{if $k = l$} \nonumber.
\end{align}
Applying  the results in the appendix to the conditional $z_{j}^{(i)} \ | \ z_{-j}^{(i)}, y_j^{(i)}$ we obtain 
\begin{align}
\label{offdiag}
E(Z^{(i)}_{j} \ | \ y^{(i)}; \widehat{ \mathbf{\Theta}}, \widehat{\mathcal{D}}) & =
 \widehat{\mathbf{\Sigma}}_{j,-j} \widehat{\mathbf{\Sigma}}_{-j,-j}^{-1} E(Z_{-j}^{(i)^t} \ | \ y^{(i)}; \widehat{ \mathbf{\Theta}}, \widehat{\mathcal{D}}) + \frac{\phi(\widehat{\delta}_{j,y_{j}^{(i)}}^{(i)}- \phi(\tilde{\delta}_{j,y_{j}^{(i)}+1}^{(i)})}{\Phi(\tilde{\delta}_{j,y_{j}^{(i)}+1}^{(i)}) - \Phi(\tilde{\delta}_{j,y_{{j}}^{(i)}}^{(i)})} \tilde{\sigma}_{j}^{(i)},
\end{align}
\begin{align}
\label{diag}
E((Z^{(i)}_{j})^2 \ | \ y^{(i)}; \widehat{ \mathbf{\Theta}}, \widehat{\mathcal{D}}) &= \widehat{\mathbf{\Sigma}}_{j,-j} \widehat{\mathbf{\Sigma}}^{-1}_{-j,-j} E(Z_{-j}^{(i)^t} Z_{-j}^{(i)} \ | \ y^{(i)}; \widehat{ \mathbf{\Theta}},\widehat{\mathcal{D}}) \widehat{\mathbf{\Sigma}}^{-1}_{-j,-j} \widehat{\mathbf{\Sigma}}^t_{j,-j}
+ (\tilde{\sigma}_{j}^{(i)})^2 \nonumber \\
	& + 2 \frac{\phi(\tilde{\delta}_{j,y_{j}^{(i)}}^{(i)})- \phi(\tilde{\delta}_{j,y_{j}^{(i)}+1}^{(i)})}{\Phi(\tilde{\delta}_{j,y_{j}^{(i)}+1}^{(i)}) - \Phi(\tilde{\delta}_{j,y_{{j}}^{(i)}}^{(i)})} [ \widehat{\mathbf{\Sigma}}_{j,-j} \widehat{\mathbf{\Sigma}}_{-j,-j}^{-1} E(Z_{-j}^{(i)^t} \ | \ y^{(i)}; \widehat{ \mathbf{\Theta}}, \widehat{\mathcal{D}})] \tilde{\sigma}_{j}^{(i)}  \nonumber \\
	& + \frac{\delta^{(i)}_{j,y_{j}^{(i)}} \phi(\tilde{\delta}_{j,y_{j}^{(i)}}^{(i)}) - \tilde{\delta}_{j,y_{j}^{(i)}+1}^{(i)} \phi(\tilde{\delta}_{j,y_{j}^{(i)}+1}^{(i)})}{\Phi(\tilde{\delta}_{j,y_{j}^{(i)}+1}^{(i)}) - \Phi(\tilde{\delta}_{j,y_{j}^{(i)}}^{(i)})} (\tilde{\sigma}_{j}^{(i)})^2,
	\end{align} 
where $Z^{(i)}_{-j} = (Z^{(i)}_{1},\ldots, Z^{(i)}_{j-1}, Z^{(i)}_{j+1},\ldots, Z^{(i)}_{p})$ and $\tilde{\delta}_{j,y^{(i)}_j}^{(i)}=[c^{(i)}_j - E(\tilde{\mu}_{ij} \ | \ y^{(i)}; \widehat{ \mathbf{\Theta}},\widehat{\mathcal{D}})] / \tilde{\sigma}_{ij}$. In this way, an approximation for $\bar{R}$ is obtained as follows:
\[ \tilde{r}_{kl}= \left\{
\begin{array}{ll}
	\frac{1}{n} \sum_{i=1}^{i=n} E(Z^{(i)}_k \ | \ y^{(i)},\widehat{ \mathbf{\Theta}}^{(m)}, \widehat{\mathcal{D}}) E(Z^{(i)}_l \ | \ y^{(i)},\widehat{ \mathbf{\Theta}}^{(m)}, \widehat{\mathcal{D}}) & \mbox{if $1 \leq k$ $\neq$ $l$ $\leq p$ }\\
	\frac{1}{n} \sum_{i=1}^{i=n} E((Z^{(i)}_k)^2 \ | \ y^{(i)}, \widehat{ \mathbf{\Theta}}^{(m)}, \widehat{\mathcal{D}}) & \mbox{if $k = l$}.
\end{array} \right. \]  

\setlength{\parindent}{0ex} \textbf{M-step.} The M-step involves updating $\mathbf{\Theta}$ by maximizing the expected complete likelihood with an $\mathcal{\ell}_1$ penalty over the precision matrix,
\[
\widehat{ \mathbf{\Theta}}_\lambda^{(m+1)}= \arg \max_ \mathbf{\Theta} \{\log |\mathbf{\Theta}| - tr(\bar{R}  \mathbf{\Theta}) - \lambda ||  \mathbf{\Theta}||_1 \}.
\]
In our implementation, we use the {\tt glasso} method for optimization \citep{witten2011new}. A multi-core implementation of our proposed methods speeds up the computational challenge, as all of the penalized optimizations are performed in parallel across the available nodes in any multi-core computer architecture. This feature proportionally reduces the computational time. Performing simulations, we noticed that the EM algorithm converges quickly, within at most $10$ iterations.

\subsection{Selection of the tuning parameter}
The penalized log-likelihood method guarantees with probability one that the precision matrix is positive definite. In addition, the method leads to a sparse estimator of the precision matrix, which encodes the latent conditional independencies between the genetic markers. Sparsistency refers to the property that all parameters that are zero are actually estimated as zero with probability tending to one. A grid of regularization parameters $\Lambda= (\lambda_1, \ldots, \lambda_N)$ controls the level of sparsity of the precision matrix. Since we are interested in graph estimation, one approach is to subsample the data, measure the instability of the edges across the subsamples \citep{liu2010stability} and to choose a $\lambda$ whose instability is less than a certain cut point value (usually taken as 0.05). However, in high dimensional settings, this approach is time consuming.

Alternatively, we compute various information criteria at EM convergence based on the observed log-likelihood, which can be written as \citep{ibrahim2008model}
\begin{equation}
\label{log-likeObs}
\ell_y(\widehat{\Theta}_\lambda) = Q(\widehat{\Theta}_\lambda | \widehat{\Theta}^{(m)}) - H (\widehat{\Theta}_\lambda | \widehat{\Theta}^{(m)}),
\end{equation}
where $Q$ is defined in (\ref{E-step}) and H function is
\[
H(\widehat{\Theta}_\lambda | \widehat{\Theta}^{(m)}_\lambda) = E_z[\ell_{Z | Y}(\widehat{\Theta}_\lambda) | Y; \widehat{\Theta}_\lambda] = E_z[\log f(z)| Y ;\widehat{\Theta}_\lambda ] - \log p(y).
\]
We consider the class of model selection criteria given by 
\begin{align}
IC_{H,Q}(\lambda)&=  -2 \ell_{z \in \mathcal{D}}(\widehat{\Theta}_\lambda) + \mbox{bias} (\widehat{\Theta}_\lambda). \nonumber
\end{align} 
Different forms of the $\mbox{bias}(\widehat{\Theta}_\lambda)$ lead to different information criteria for model selection. As we are interested in graph estimation, we use the extended Bayesian information criterion (eBIC) introduced for conditional independence graph selection \citep{foygel2010extended}
\[
eBIC(\lambda) = -2 \ell(\widehat{\Theta}_\lambda) +  ( \log n + 4 \gamma \log p) df(\lambda),
\]
where $df(\lambda)= \sum_ {1 \leq i < j \leq p} I(\widehat{\theta}_{ij, \lambda} \neq 0 )$ refers to the number of non-zero off-diagonal elements of $\widehat{\Theta}_\lambda$ and $\gamma \in [0,1]$ is the parameter that penalizes the number of models, which increases when $p$ increases. In case of $\gamma = 0 $ classical BIC is obtained. Typical values for $\gamma$ are 1/2 and 1. To obtain the optimal model in terms of graph estimation we pick the penalty term that minimizes EBIC over $\lambda > 0$. 
\begin{figure} 
\centering{
\includegraphics[width=0.5\textwidth]{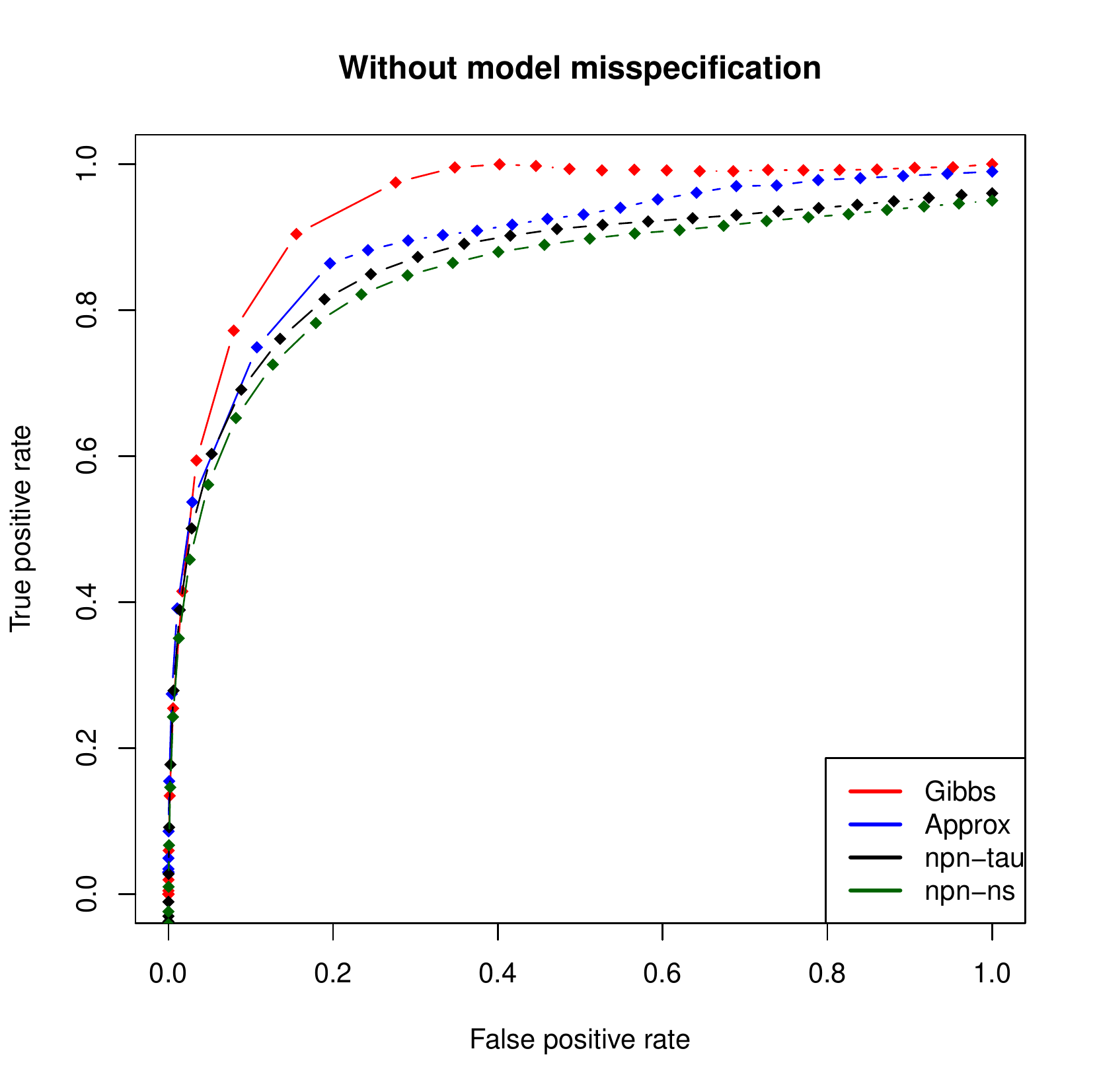}%
\includegraphics[width=0.5\textwidth]{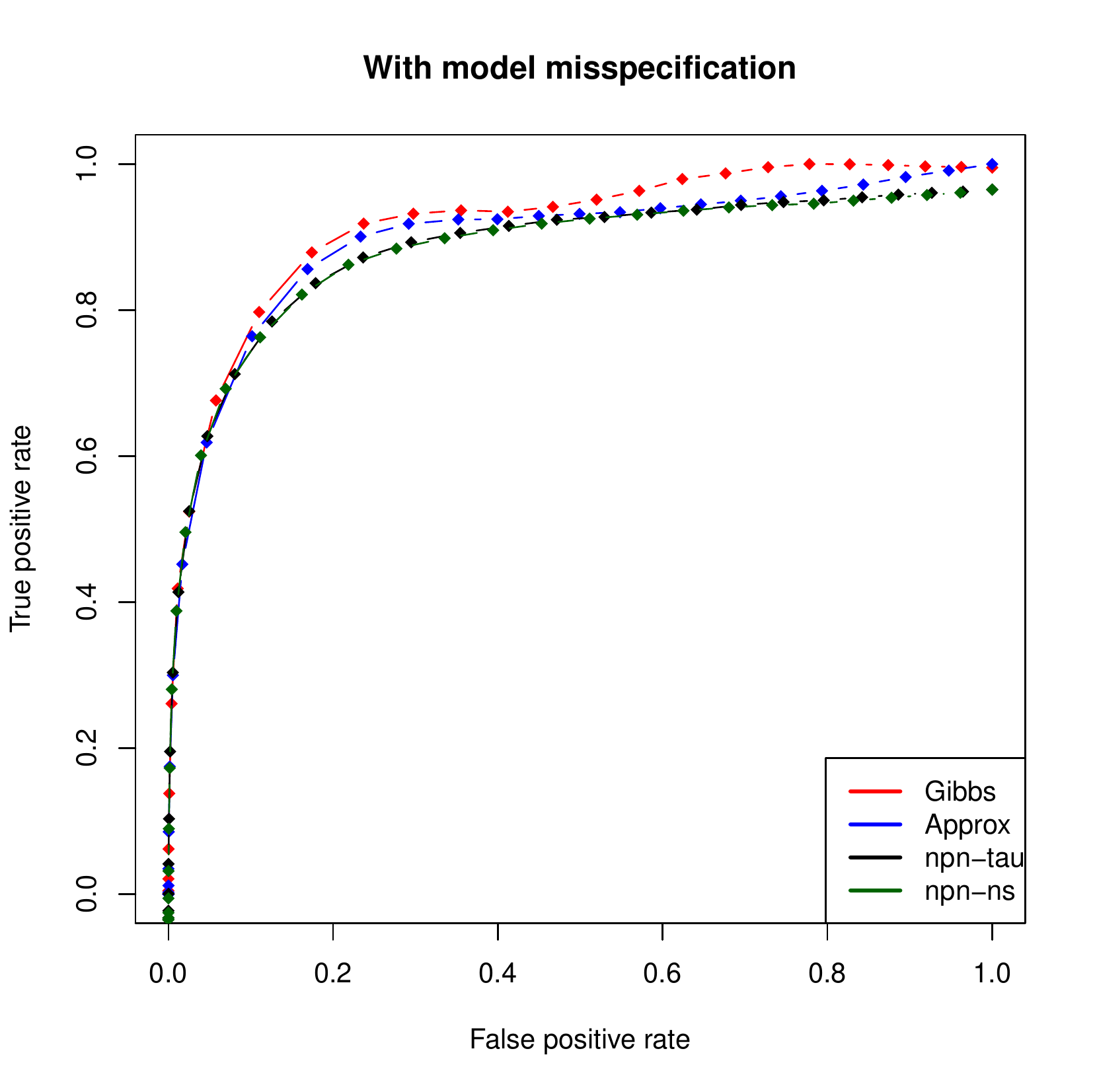}%
}
 
(a) \hspace{5.15cm} (b)
\caption{\scriptsize ROC curves for comparing different methods of recovering the true graph where $p=1000$, $n=100$ and, $k=3$. The data is simulated from (a) our the Gaussian copula graphical model, and (b) $t_{(3)}$ copula graphical model. Our method (Gibbs) consistency outperforms other methods.}
\label{ROC}
\end{figure}

\subsection{Inference uncertainty}
\label{uncertainty}

The classical likelihood-based method to estimate uncertainty by inverting the Fisher information matrix does not directly apply to penalized likelihood approaches \citep{lehmann1998theory}. Instead, one way to compute uncertainty associated with the estimation of precision matrix under the penalized Gaussian copula graphical model is through a non-parametric bootstrap. For the penalized likelihood bootstrap, we replicate $B$ datasets that are created by sampling with replacement $n$ samples from the dataset $Y_{n \times p}$. We treat each replicate as the original data and run the entire inference procedure of the proposed Gaussian copula graphical model to estimate $\widetilde{\Theta}^{(b)}_{\hat{\lambda}}$ ($b=1,\ldots,B$). In this bootstrap, we take into account the uncertainty arising from both empirical estimation of marginals and selection of the tuning parameter. Thus, the above mentioned non-parametric bootstrap procedure adequately reflects the underlying uncertainty in the estimation procedure of the proposed epistatic interaction graph. We have implemented this procedure to evaluate the uncertainty associated with the estimation of the epistatic interactions in the Arabidopsis thaliana experiment in section \ref{bootstrapThaliana}.
\begin{table} 
\caption{\scriptsize The comparison between the performance of the proposed regularized approximated EM, regularized Gibbs sampler EM, the nonparanormal skeptic Kendall's tau, and the nonparanormal normal-score. The means of the $F_1$-score, sensitivity and specificity over $75$ replications are represented. The high value of the $F_1$-score is the indicator of good performance. The best model in each column is boldfaced.}
\centering
{\scriptsize	
\begin{tabular}{l*{15}{l}l}
&\multicolumn{3}{c}{p=90, n=360, k=3} & \multicolumn{3}{c}{p=1000, n=200, k=3 }   \\ 
     \cmidrule(l{1pt}r{1pt}){3-4}   \cmidrule(l{2pt}r{2pt}){5-7} 
&& Normal& \quad $t_{(3)}$&& Normal& \quad$t_{(3)}$ \\
\midrule
\multicolumn{2}{l}{\textbf{Gibbs}}  \\
$F_1$ oracle && 0.83(0.02)& 0.83(0.02)&& 0.75(0.04)& 0.76 (0.02) \\
$F_1$&& \textbf{0.76}(0.03)& \textbf{0.75}(0.03)&& \textbf{0.74}(0.04)& \textbf{0.50} (0.06) \\
SEN  && 0.97(0.02)& 0.98(0.01)&& 0.67(0.07)& 0.26 (0.05) \\
SPE  && 0.97(0.00)& 0.97(0.00)&& 0.99(0.00)& 0.99 (0.00) \\
\\
\multicolumn{2}{l}{\textbf{Approx}}    	\\ 
$F_1$ oracle && 0.80(0.02)& 0.80(0.02)&& 0.73(0.03)& 0.74 (0.02) \\
$F_1$&& 0.70(0.03)& 0.70(0.03)&& \textbf{0.73}(0.03)& \textbf{0.50} (0.35) \\
SEN  && 0.98(0.02)& 0.96(0.01)&& 0.70(0.08)& 0.50 (0.35) \\
SPE  && 0.96(0.01)& 0.98(0.01)&& 1.00(0.00)& 1.00(0.00) \\
\\
\multicolumn{2}{l}{\textbf{npn-tau}}    	\\ 
$F_1$ oracle && 0.84(0.02)& 0.84(0.02)&& 0.76(0.03)& 0.76 (0.02) \\
$F_1$&& 0.70(0.15)& 0.70(0.15)&& 0.00(0.00)& 0.00 (0.00) \\
SEN  && 0.94(0.19)& 0.94(0.19)&& 0.00(0.00)& 0.00 (0.00) \\
SPE  && 0.97(0.01)& 0.97(0.01)&& 1.00(0.00)& 1.00 (0.00) \\
\\
\multicolumn{2}{l}{\textbf{npn-ns}}    	\\ 
$F_1$ oracle && 0.83(0.02)& 0.83(0.02)&& 0.75(0.03)& 0.75 (0.03) \\
$F_1$&& 0.65(0.25)& 0.56(0.32)&& 0.00(0.00)& 0.00 (0.00) \\
SEN  && 0.86(0.32)& 0.74(0.42)&& 0.00(0.00)& 0.00 (0.00) \\
SPE  && 0.97(0.01)& 0.98(0.01)&& 1.00(0.00)& 1.00 (0.00) \\
\bottomrule
\end{tabular} }
\label{tableCom}
\end{table}

\section{Simulation study}
\label{simulation} 
We study the performance of the proposed method in a simulation study, mimicking the small-sized genotyping study involving Arabidopsis and the medium-sized study involving maize. For each dimension, we consider two different scenarios: in one scenario the latent variables satisfy the multivariate Gaussian distribution, and in the other scenario they do not. In the latter, we consider the t-distribution with a $3$ degrees of freedom. The simulated data are obtained by different scenarios for the number of variables $p \in \{90, 1000\}$, the number of sample sizes $n \in \{200, 360\}$, and a fixed genotype state $k=3$. 

The simulated graphs mimic a true underlying epistasis selection network. First, we partition the variables into $g$ linkage groups (each of which represents a chromosome), then within each linkage group adjacent markers are linked via an edge due to genetic linkage. Also, with probability $\alpha= 0.01$ a pair of non-adjacent markers in the same chromosome is given an edge. Trans-chromosomal edges are simulated with probability $\beta= 0.02$. In the low-dimension case $(p=90)$ we created $5$ chromosomes, and in high-dimension case $(p =1000)$ $10$ chromosomes. The corresponding positive definite precision matrix $\Theta$ has a zero pattern corresponding to the non-present edges. For each iteration in the simulations a new random precision matrix was generated. The latent variables are simulated from either a multivariate normal distribution, $N_p (0, \Theta^{-1})$, or a multivariate t-distribution with $3$ degrees of freedom and covariance matrix $\Theta^{-1}$. We generate random cutoff points from the uniform distribution. And we discretize the latent space into $k= 3$ disjoint states.  

We compare our proposed method with other approaches in terms of ROC performance. Also, we compare our model to other methods in terms of graph recovery.

The ROC curves in Figure \ref{ROC} show the performance of the different graph estimation methods. The area under the curve is used as a measure of the quality of the methods in recovering the true graph. Here, we compare the following methods:
\begin{enumerate}
\item Our method with Gibbs sampler within the EM, (Gibbs).
\item Proposed method with approximation within the EM, (Approx)
\item Nonparanormal skeptic using Kendall's tau, (NPN-tau) \citep{liu2012high}
\item Nonparanormal normal-score, (NPN-ns) \citep{liu2009nonparanormal}
\end{enumerate}
Figure \ref{ROC} shows the average false positive and true positive rates over $75$ repeated simulations each at $30$ grid points of the tuning parameter. In Figure \ref{ROC}(a), the latent variable satisfy the Gaussian distribution, and (b) the latent variable is non-Gaussian. Figure \ref{ROC} shows how our proposed method, particularly the one employing the Gibbs sampler, outperforms the nonparanormal approaches. This is true both in the scenario of no model misspecification, i.e., that the data is simulated from our the Gaussian copula graphical model, as in the case of model misspecification, i.e., when the data is simulated from the $t_{(3)}$ copula graphical model. Our method combined with the approximation approach performs somewhat better than both nonparanormal approaches in both under the true model and in the case of model misspecification. Based on our simulation studies the performance of the NPN-tau and the NPN-ns are similar in the absence of outliers, as discussed in \cite{liu2012high}.
 
Furthermore, we measure the methods' performance in terms of graph accuracy and its closeness to the true graph. The above penalized inference methods are a path-estimation procedures, however, in practice, a particular network should be selected. As we are interested in the global true interaction structure, but not in the individual parameters, the extended Bayesian Information Criterion (eBIC) is particularly appropriate. To evaluate the accuracy of the estimated graph we compute the $F_1$-score $(F_1\mbox{-score} = \frac{2 {TP}}{2 TP + FP + FN})$, sensitivity $(\mbox{SEN} = \frac{TP}{TP + FN})$ and specificity $(\mbox{SPE} = \frac{TN}{TN + FP})$, where TP, TN, FP, FN are the numbers of true positive, true negative, false positive and false negative values, respectively, in identifying the nonzero elements in the precision matrix. We note that high values of the $F_1$-score, sensitivity and specificity indicate good performance of the proposed approach for the given combination of $p$, $n$ and $k$. However, as there is a natural trade-off between sensitivity and specificity, we focus particularly on the $F_1$ score to evaluate the performance of each method. For each simulated dataset, we apply each of the four methods. 

In Table \ref{tableCom}, we compare these four methods in a low-dimensional case $p=90$, $n=360$, and $k=3$ mimicking the Arabidopsis dataset we consider later, and a high-dimensional case of $p=1000$, $n=200$, and $k=3$, mimicking the Maize genotype data. In both cases we consider two different scenarios: in one scenario the latent variable satisfies the Gaussian distribution, and in the other scenario it is overdispersed according to a $t_{(3)}$. We report the average values for $F1$-score, SEN and SPE in $75$ independent simulations. The value of $F_1$ oracle indicates the best values of $F_1$ that can be obtained by selecting the best tuning parameter in the $\ell_1$ optimization. Table \ref{tableCom} shows that the proposed method either using the Gibbs sampling or the approximation method within the EM performs very well in selecting the best graph. In both scenarios in the low dimension case, the NPN-tau chooses a better graph compared to NPN-ns. However, in the high-dimension case neither of them chooses a best graph. In fact, they select an empty graph. The other measurement, namely sensitivity indicates the true edges that we recover in the inferred network. The high value of specificity shows that the zero entries in the precision matrix, i.e., the absent edges in the network are accurately identified. These results suggest that, though recovering sparse network structure from discrete data is a challenging task, the proposed approaches perform well. 

\begin{table} 
\centering
\caption{
\scriptsize The computational cost comparison (in minutes) between the proposed method (Approx) and the nonparanormal skeptic (NPNtau) method. For the larger $p$'s the nonparanormal skeptic method is faster than our proposed method. However, neither the npn-tau nor npn-ns can deal with the missing values, while the proposed approximation approach is developed to be able to deal with missing genotypes that commonly occur in genotype data. \label{timeCost} }

{\scriptsize
  \begin{tabular}{ccccccc}
  \addlinespace
  \toprule
  {\bf } & \multicolumn{ 6}{c}{{ Number of variables}} \\
  & { 100} & { 500} & { 1000} & { 2000} & { 3000} & { 4000} \\
  \hline
  Approx   & 0.34  & 1.26 & 19.71  & 80.43  & 734.79  & 2623.68 \\
  npn-tau  & 0.03  & 0.16 & 1.76  & 14.05 & 62.76     & \textendash$^*$   \\
  \bottomrule
  \end{tabular}}
\vspace{.2cm} \\
\scriptsize {$^*$ Exceeded step memory limit at some point}
\end{table}

We perform all the computations on a cluster with $24$ Intel Xeon $2.5$ GHz cores processor and $128$ GB RAM. In our proposed method it is possible to estimate the conditional expectation either through Gibbs sampling, or the approximation approach. For large numbers of markers ($p \ge 2000$) the Gibbs sampling approach is not recommended due to excessive computational costs. However, the approximation approach is able to handle such situations. The computational costs for the non-paranormal skeptic and the normal-score methods are similar to each other. Thus, in Table \ref{timeCost} we report the computational cost of the proposed approximation method and the non-paranormal skeptic method versus the number of variables, for a sample size fixed to $200$. Both methods have a roughly similar increase in computational time, which seems to be larger than quadratic in $p$. Our method is roughly a constant factor $10$ slower than the nonparanormal skeptic. This is due to the EM iterations. The EM has advantages, however, as our method is able to deal with missing genotype values, which are very common in practice. However, by programming in multi-core we have significantly reduced the computational costs. Further improvement will be achieved by programming in \textbf{C++} and interfacing it with \textbf{R}.
\begin{figure}[t]
\centering{
\includegraphics[width=0.45\textwidth]{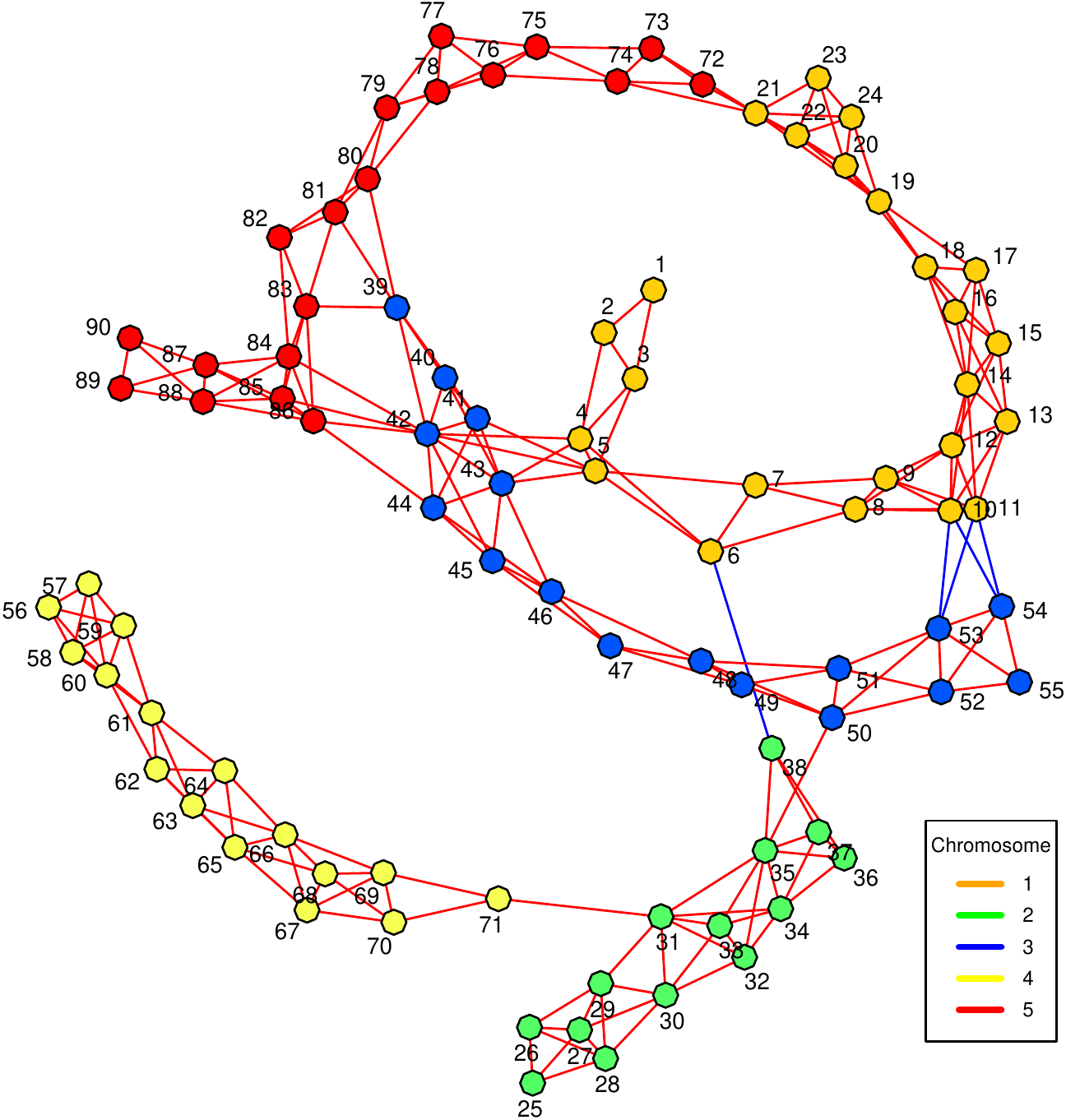}%
\hspace{0.4cm}
\includegraphics[width=0.5\textwidth]{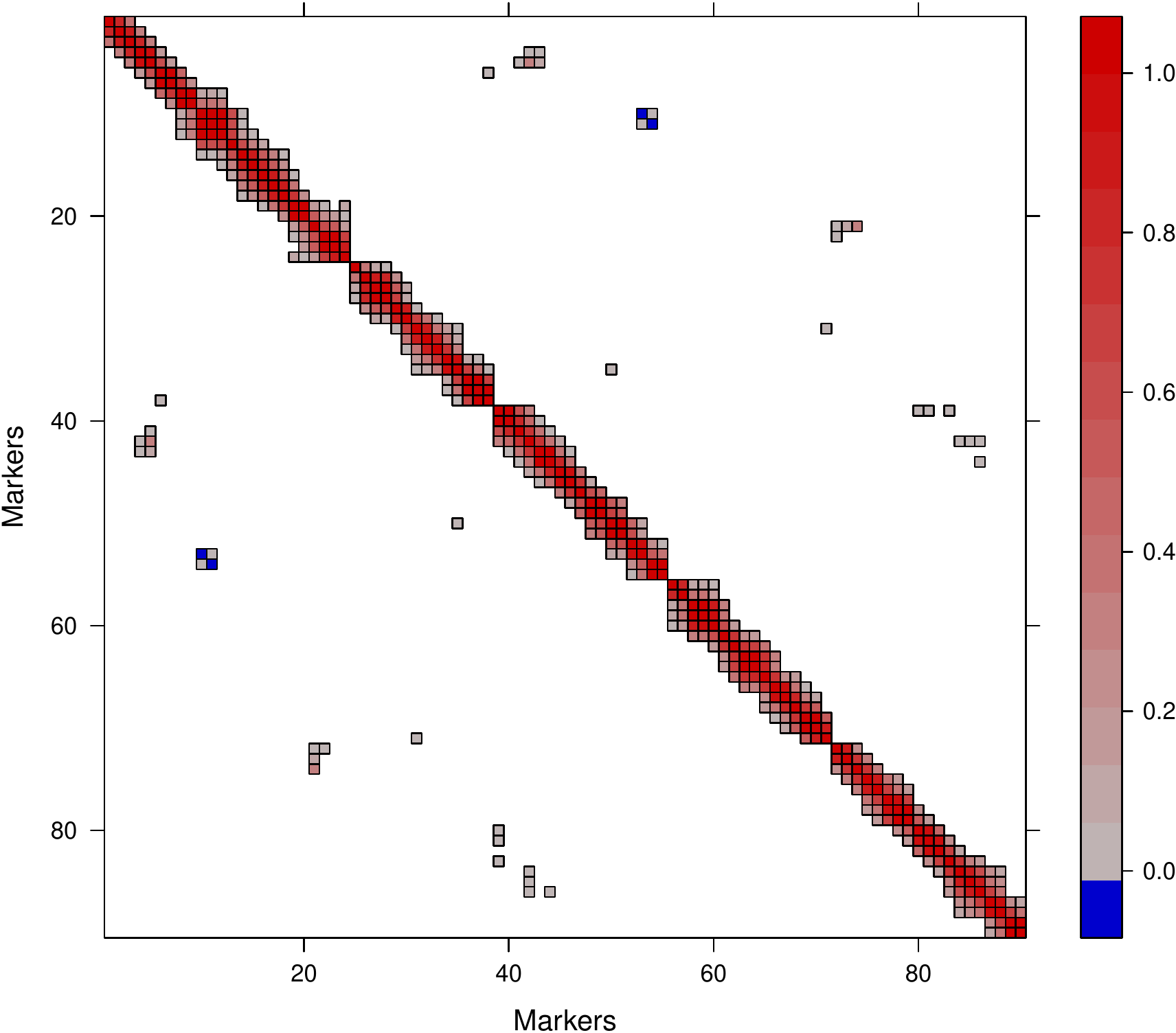}%
}
\hspace{5cm}(a) \hspace{6cm} (b)
\caption{\scriptsize The inferred network for the genotype data in cross between the $A.thaliana$ accessions, Columbia-0 (col-0) and the Cape Verde Island (Cvi-0). (a) Each color corresponds to different chromosomes in \emph{A. thaliana}. Nodes (genetic markers) with similar colors belong to the same chromosome. The different edge colors show the positive (red) and negative (blue) partial correlations. (b) represents the zero pattern of the partial correlation matrix.}
\label{cvicol}
\end{figure}

\section{Detecting genomic signatures of epistatic selection}
\subsection{Epistatic selection in \emph{Arabidopsis thaliana}}
\label{arabi}
We apply our proposed Gibbs sampling approach to detect epistatic selection in \emph{Arabidopsis thaliana} genotype data that are derived from a RIL cross between Columbia-0 (Col-0) and the Cape Verde Island (Cvi-0), where $367$ individual plants were genotyped across $90$ genetic markers \citep{simon2008qtl}. The $Cvi-0 \times Col-0$ RIL is a diploid population with three possible genotypes, $k = 3$. The genotype data are coded as $\{0, 1, 2\}$, where $0$ and $2$ represent two homozygous genotypes (AA resp. BB) from Col-0 and Cvi-0, $1$ defines the heterozygous genotype (AB). Some markers have missing genotypes $(0.2\%)$.

The results of the analysis are presented in Figure \ref{cvicol}. The first thing to note is that the Gaussian copula graphical model groups together markers that belong to one chromosome, because of genetic linkage. In the diagonal of Figure \ref{cvicol}(b), the $5$ chromosomes of the Arabidopsis are clearly identifiable. If there is no linkage disequilibrium, markers in different chromosomes should segregate independently; in other words, there should be no conditional dependence relationships between markers in different chromosomes. Existence of trans-chromosomal conditional dependencies reveal the genomic signatures of epistatic selection. Figure \ref{cvicol} shows that our method finds some trans-chromosomal regions that do interact. In particular, the bottom of chromosome $1$ and the top of chromosome $5$ do not segregate independently of each other. Beside this, interactions between the tops of chromosomes $1$ and $3$ involve pairs of loci that also do not segregate independently. This genotype has been studied extensively in \cite{bikard2009divergent}. They reported that the first interaction we found causes arrested embryo development, resulting in seed abortion, whereas the latter interaction causes root growth impairment. 

Furthermore, in addition to these two regions, we have discovered a few other trans-chromosomal interactions in the \emph{Arabidopsis thaliana} genome. In particular, two adjacent markers, $c1-13869$ and $c1-13926$, in the middle of the chromosome $1$ interact epistatically with the adjacent markers, $c3-18180$ and $c3-20729$, at the bottom of chromosome $3$. The sign of their conditional correlation score is negative indicating strong negative epistatic selection during inbreeding. These markers therefore seem evolutionarily favored to come from different grandparents. This suggests some positive effect of the interbreeding of the two parental lines: it could be that the paternal-maternal combination at these two loci protects against some underlying disorder or that it actively enhances the fitness of the resulting progeny.

\begin{table}
\centering
\caption{ 
{\scriptsize A summary of model fit to the Arabidopsis genotype data.\label{ShrunkmodelFit}}
} 

{\scriptsize
\begin{tabular}{l*{8}{c}}
\hline
Model &   df & Log-likelihood & Deviance & P-value \\
\hline
Fitted model    & 	$237$ &  $-1098.75$  \\
Saturated model &  $4005$ & $193.35$ \\ 
Fitted model vs Saturated model & $3768$ &	&	$2584.2$  & $1$   \\
\hline
\end{tabular}}
\end{table}
\subsubsection{Fit of model to A.thaliana data}
\label{Goodness of fit}
Calculating the deviance statistics $D$ allows us to assess the goodness-of-fit of the proposed method,
\[
D = -2 [\ell_m(\widehat{\Theta}) - \ell_s(\widehat{\Theta}) ],
\]
where $\ell_m(\widehat{\Theta})$ and $\ell_s(\widehat{\Theta})$ denote the log-likelihood of the observations for the fitted model and the saturated model, respectively.

In our modeling framework, the log-likelihood of the fitted model corresponds to the $\ell_Y(\widehat{\Theta}_\lambda)$ that we obtain from the equation (\ref{log-likeObs}). Taking out the penalty term from (\ref{likelihood}) we obtain the non-penalized log-likelihood of the saturated model, as follows:
\[
\ell_s(\widehat{\Theta}) = \ell_Y(\bar{R}) \cong - \frac{n}{2} \log |\bar{R}| - \frac{1}{2} n p,
\]
where $\bar{R}$ is the estimated covariance matrix that can be calculated through either Gibbs sampling or approximation approaches in sections \ref{inference} A or B.

Table \ref{ShrunkmodelFit} shows how well the proposed model fits the A.thaliana data. The $\chi^2$ test with $3768$ degrees of freedom gives a p-value of $1$, indicating that the proposed model fits the data adequately.

\subsubsection{Evaluating the estimated network in A.thaliana}
\label{bootstrapThaliana}
\begin{figure}[t!] 
{\hspace{-0.8cm}
\includegraphics[width=6.5in,angle=-90]{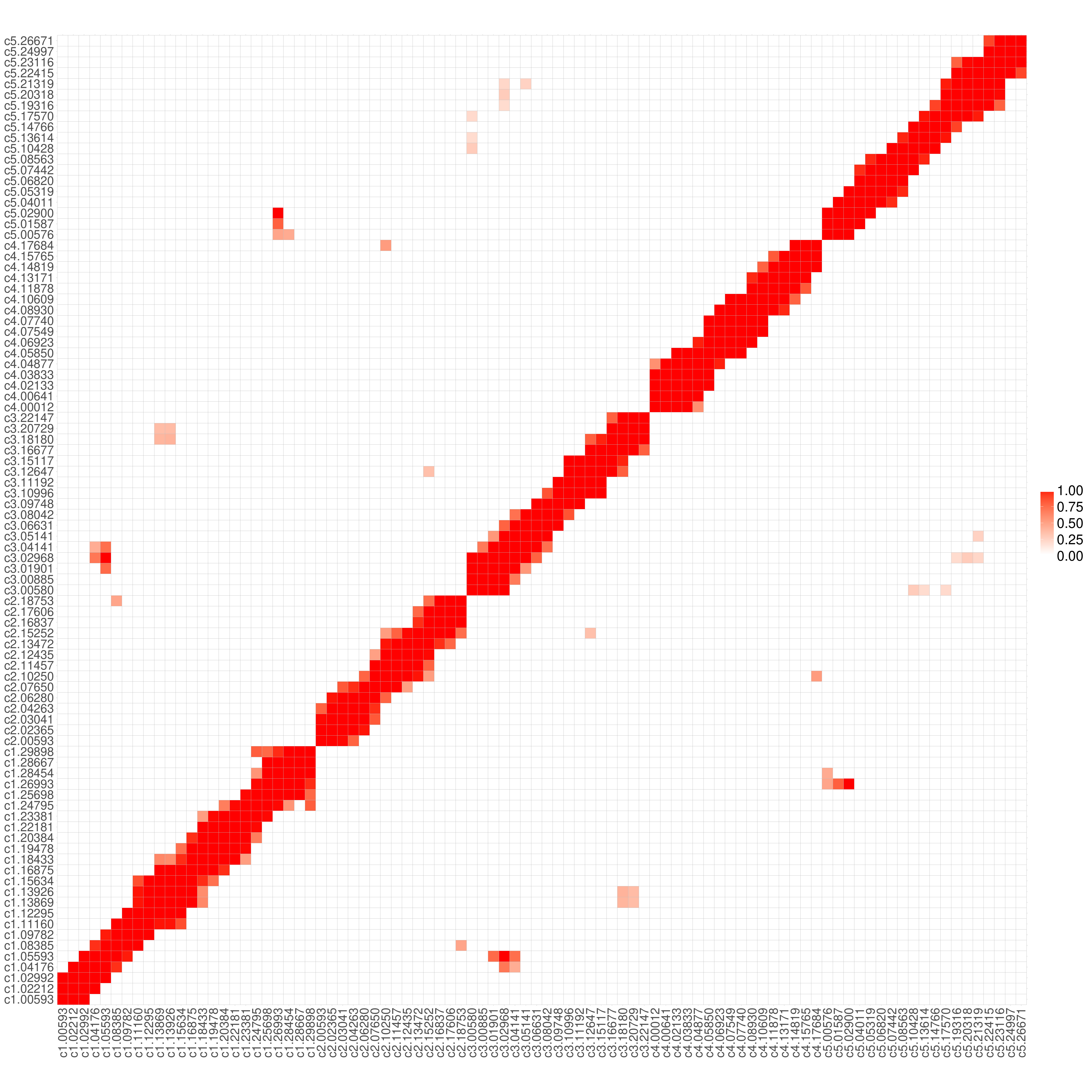}%
}
\caption{\scriptsize The uncertainty associated with the estimation of the precision matrix in \emph{A.thaliana} using the non-parametric bootstrap. The off-diagonal elements represent the probability of having positive or negative epistatic interactions between markers in different chromosomes in bootstrap versions of the data, whereas the ``thick'' diagonal elements show the relative frequency of having links between neighboring markers within the chromosomes in the bootstrapped data.}
\label{probExistingLinks}
\end{figure}

We use a non-parametric bootstrapping approach to determine the uncertainty associated with the estimated edges in the precision matrix in \emph{A.thaliana}. We generate $200$ independent bootstrap samples --- as described in section \ref{uncertainty} --- from the genotype data of Col-0 and Cvi-0 cross. For each $200$ bootstrap samples, we apply the proposed Gaussian copula graphical model as described in section \ref{GMEpi}. Furthermore, we calculate the frequency of each entry in $\tilde{\Theta}^{b}_{\hat{\lambda}}$ ($b=1,\ldots, 200$) have the same sign as the estimated $\widehat{\Theta}_{\hat{\lambda}}$ from the original Cvi-0 and Col-0 genotype data. In Figure \ref{probExistingLinks}, we report the corresponding relative frequencies for a sign match of each link across the $200$ bootstrap samples.

Figure \ref{probExistingLinks} shows the uncertainty associated with the epistatic interactions between markers in chromosomes 1 and 5. In particular, in all bootstrap samples we infer a positive epistatic interaction between markers c1-26993 and c5-02900. Also their neighboring markers interact epistatically. Another region in the \emph{A.thaliana} genome that contains epistatic interactions is between chromosomes 1 and 3. In all bootstrap samples, we infer positive epistatic interaction between markers c1-05593 and c3-02968, including their neighboring markers. \cite{bikard2009divergent} show that these two regions cause arrested embryo development and root growth impairment in \emph{A.thaliana}, respectively. In addition to these two confirmed regions, we have found other trans-chromosomal regions with potential epistatic interactions. For example, two neighboring markers in chromosomes 1, namely c1-138669 and c1-13869, have quite consistent negative epistatic interactions with the two neighboring markers in chromosome 3, namely c3-20729 and c3-18180.  

\begin{figure}[t]
\centering
\includegraphics[width=0.7\textwidth]{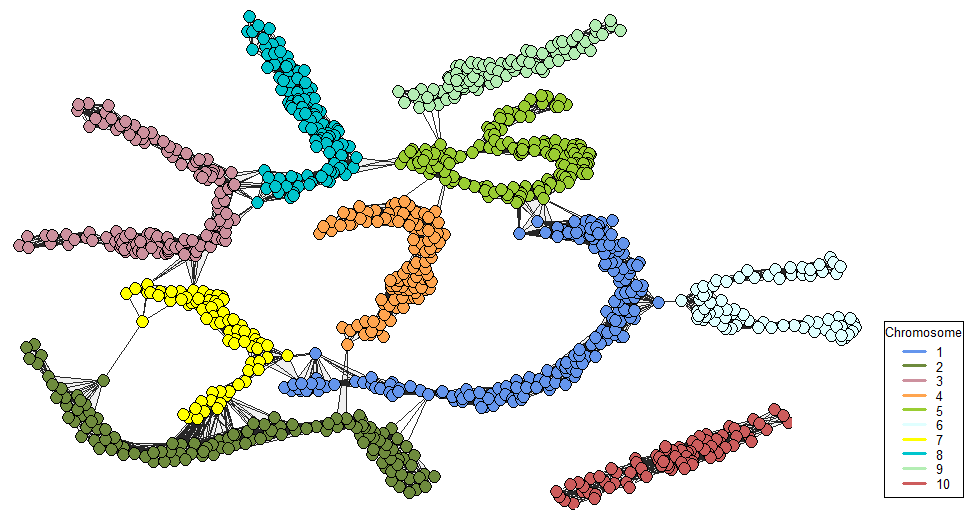} 
\caption{\scriptsize The inferred network for $1106$ markers in the cross between $B73$ and $Ki11$ in maize using approximated method in Gaussian copula graphical model.}
\label{maizenet}
\end{figure}
\subsection{Genetic inbreeding experiment in maize}
\label{maize}
The Nested Association Mapping (NAM) initiative in maize populations is designed to reveal the genetic structure of underlying complex traits in maize \citep{mcmullen2009genetic, rodgers2015recombination}. As part of this study, an inbred $Ki11$ maize line was crossed with the $B73$ reference line.  This genotype data contains $1106$ markers  genotyped for $193$ individuals. The $B73 \times Ki11$ RIL is a diploid population with three possible genotypes, $k = 3$. We applied our proposed approximation method to the $B73 \times Ki11$ sample, aiming to reveal genetic regions in the maize genome that interact epistatically and may lead to maize disease, e.g. growth impairments, lower fertility or sterility. Exploring genomic signatures of such high-dimensional epistatic selection has so far been left unexplored in previous analyses of this essential crop. Figure \ref{maizenet} shows that certain loci on different chromosomes do not segregate independently of each other. For instance, marker $PZA02117.1$ in chromosome $1$ interacts with markers $PZA02148.1$ in chromosome $6$, and marker $PZA00545.26$ in chromosome $5$ interacts with three adjacent markers $PZA00466.1$, $PZA01386.3$, and $PZA02344.1$ in chromosome $9$. Existence of such trans-chromosomal conditional dependencies indicates marker-marker associations that are possibly due to epistatic selection. Statistically speaking, conditional dependence relationships between physically unlinked pairs of genetic regions contribute to some disorders in this crop that affect its viability.

\section{Discussion}
\label{discussion}
Epistatic selection involves the simultaneous selection of combinations of genotypes at two or more loci. Epistatic selection can create linkage disequilibrium (LD) between loci, within and across chromosomes. These LD distortions point to genomic regions undergoing selection. Epistasis is widespread but it may often go undetected due to lack of statistical power due to testing multiple hypotheses in a possibly very high-dimensional setting, experimental challenges due to the large sample sizes that are required in order to detect significant interactions, and computational challenges which relate to dealing with missing genotypes and the large number of tests to be evaluated. 

In this paper, we have introduced an efficient alternative method based on Gaussian copula graphical models that models the phenomenon of epistasis sparsely in a high-dimensional setting. It is important to remember that this model is the simplest possible multivariate ordinal model as it uses the least number of parameters --- $\frac{p(p-1)}{2}$ as $\Theta$ is symmetric and the diagonal of $\Theta^{-1}$ is constrained to be 1 --- to describe the full multivariate dependence structure.  The proposed method can handle missing genotype values, and it captures the conditional dependent short- and long-range LD structure of genomes and thus reveals {\textquotedblleft aberrant\textquotedblright} marker-marker associations that are due to epistatic selection rather than gametic linkage. Polygenic selection on loci that act additively can easily be detected on the basis of strong allele-frequency distortions at individual loci. Epistatic selection, by contrast, does not produce strong locus-specific distortion effects but instead leads to pair-wise allele frequency changes. 

The proposed method explores the conditional dependencies among large numbers of genetic loci in the genome. To obtain a sparse representation of the high-dimensional genetic epistatic network, we implement an $\ell_1$ penalized likelihood approach. Other extensions of Gaussian graphical models have also been proposed.  \cite{vogel2011elliptical} extend Gaussian graphical models to elliptical graphical models, whereas \cite{finegold2009robust} provide a latent variable interpretation of the generalized partial correlation graph for multivariate t-distributions. They also employ an EM-type algorithm for fitting the model to high dimensional data.

In the application of our method to a Arabidopsis RIL, we discovered two regions that interact epistatically, which had prior been shown to cause arrested embryo development and root growth impairments. In addition, we employed our method to reveal genomic regions in maize that also do not segregate independently and may lead to lower fertility, sterility, complete lethality or other maize diseases. Although Arabidopsis thaliana and Maize are both diploid species, nothing in our method is limited to diploids. For triploid species, such as seedless watermelons, or even decaploid species, such as certain strawberries, the method can be used to find epistatic selection by merely adjusting the parameter $k$ (from 3 to, respectively, 4 and 11).



\section*{Appendix}
\label{app} 
The following results on the conditional first and second moment of the truncated normal are used in  (\ref{offdiag}) and (\ref{diag}). Suppose a random variable $X$ follows a Gaussian distribution with mean $\mu_0$ and variance $\sigma_0$. For any constant $t_1$ and $t_2$, $X \vert  t_1 \leq X \leq t_2$ follows a truncated Gaussian distribution defined on $[t_1, t_2]$. Let $\epsilon_1 = (t_1-\mu_0)/\sigma_0$ and $\epsilon_2 = (t_2 - \mu_0)/ \sigma_0$, then the first and second moments are
\begin{equation}
E(X | t_1 \leq X \leq t_2) = \mu_0 + \frac{\phi(\epsilon_1) - \phi(\epsilon_2)}{\Phi(\epsilon_2) - \Phi(\epsilon_1)} \sigma_0 \nonumber
\end{equation}
\begin{equation}
E(X^2 | t_1 \leq X \leq t_2)= \mu_0^2 + \sigma_0^2 + 2 \frac{\phi(\epsilon_1) - \phi(\epsilon_2)}{\Phi(\epsilon_2) - \Phi(\epsilon_1)} \mu_0 \sigma + \frac{\epsilon_1 \phi(\epsilon_1) - \epsilon_2 \phi(\epsilon_2)}{\Phi(\epsilon_2) - \Phi(\epsilon_1)} \sigma_0^2 \nonumber
\end{equation}
where $\Phi^{-1}$ defines the inverse function of CDF of standard normal distribution.

\section*{Acknowledgements}
The authors would like to thank Danny Arends for his helpful suggestions with respect to the software implementation of the method.\vspace*{-8pt}

\section{Supplementary Materials}
\label{supplementaryMaterials}

In the Gibbs sampler, all samples can be accepted except those obtained from a burn-in period. To find how many iterations need to be discarded (burn-in samples) we perform the Heidelberger-Welch test \citep{heidelberger1983simulation}. In this convergence test the Cramer-von-Mises statistic is used to test the null hypothesis that the sampled values derived from a stationary distribution. First the test is performed on the whole chain, then on the first $10\%, 20\%, \ldots$ of the chain until either the null hypothesis is accepted, or $50\%$ of the chain has been discarded. The output of the test is either a failure, meaning that longer iteration is needed, or a pass which then reports the number of iterations that needed to be discarded.  

\begin{figure}[t]
\centering
\includegraphics[width=0.4\textwidth]{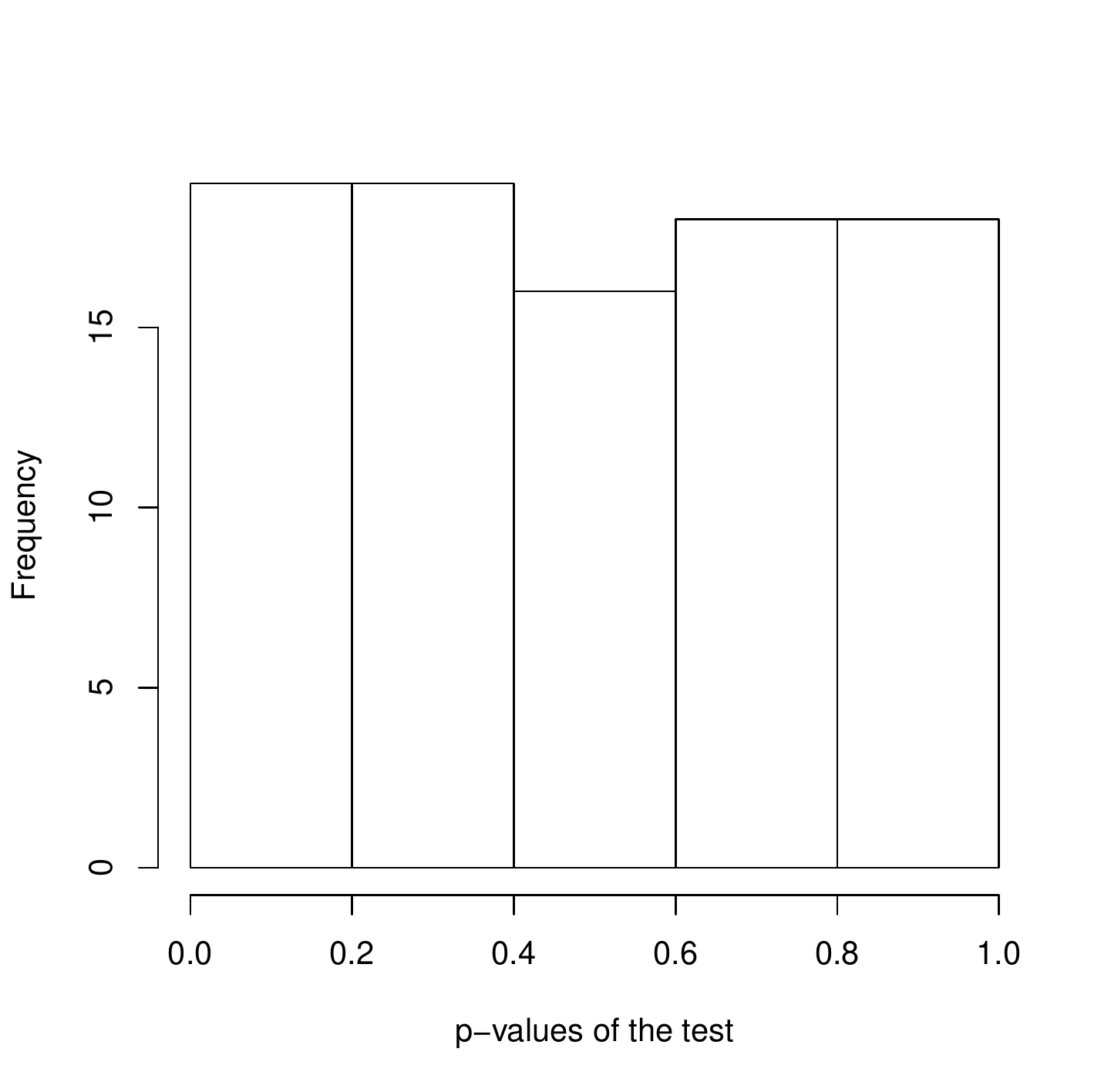}
\caption{Randomness of the derived p-values from performing the Heidelberger-Welch test in the Col $\times$ Cvi genotype data. \label{hist_pvalues}}
\end{figure}

In the implementation of the method, we took $1000$ burn-in samples in order to generate randomly from the truncated normal distribution. We show here that this amount of burn-in period is more than sufficient. As an example we apply this test in the Arabidopsis genotype data to check the number of needed burn-in samples to calculate the conditional expectation. Table \ref{Heidelberger-Welch test} shows that many variables need only one iteration to pass the stationary test. Only very few variables need either $151$ or $101$ iterations to pass the stationary test. Furthermore, we compare the derived p-values from the test with the randomly generated values from the uniform distribution (see Figure \ref{hist_pvalues}). From Table \ref{Heidelberger-Welch test} and Figure \ref{hist_pvalues}, we conclude that we do not need many burn-in samples; fixing it to $1000$ is sufficient.

\begin{table} 
\scriptsize
\caption{ A Heidelberger-Welch test to determine the needed number of burn-in samples in the Gibbs sampler within EM copula in the Col $\times$ Cvi genotype data. The start iteration shows the number of iterations that are needed to pass the stationary test.}
\label{Heidelberger-Welch test}
{\scriptsize
\begin{minipage}{0.45\textwidth}
\centering
\begin{tabular}{l*{10}{c}}
 Marker  & stationarity  & start   &   p-value \\
 & test & iteration & \\
\hline                    
$c1.00593$ & passed    &     1		  & 0.4113 \\
$c1.02212$ & passed    &     1 	  & 0.3822  \\
$c1.02992$  & passed    &     1 	  & 0.8782  \\
$c1.04176$  & passed    &     1	  & 0.3080  \\
$c1.05593$  & passed    &     1 	  & 0.1100 \\
$c1.08385$  & passed    &     1 	  & 0.1393  \\
$c1.09782$  & passed    &     1       & 0.6770 \\
$c1.11160$  & passed    &     1       & 0.6771 \\
$c1.12295$  & passed    &     1       & 0.8242 \\
$c1.13869$ & passed    &     1       & 0.5021 \\
$c1.13926$ & passed    &     1       & 0.1504 \\
$c1.15634$ & passed    &     1       & 0.1296 \\
$c1.16875$ & passed    &     1       & 0.7198 \\
$c1.18433$ & passed    &     1       & 0.3059 \\
$c1.19478$ & passed    &     1       & 0.7271 \\
$c1.20384$ & passed    &     1       & 0.8084 \\
$c1.22181$ & passed    &     1       & 0.7927 \\
$c1.23381$ & passed    &     1       & 0.6483 \\
$c1.24795$ & passed    &     1       & 0.7603 \\
$c1.25698$ & passed    &     1       & 0.7328 \\
$c1.26993$ & passed    &     1       & 0.1750 \\
$c1.28454$ & passed    &     1       & 0.7736 \\
$c1.28667$ & passed    &     1       & 0.2545 \\
$c1.29898$ & passed    &     1       & 0.9053 \\
$c2.00593$ & passed    &     1       & 0.8241 \\
$c2.02365$ & passed    &     1       & 0.8459 \\
$c2.03041$ & passed    &     1       & 0.8789 \\
$c2.04263$ & passed    &     1       & 0.6825 \\
$c2.06280$ & passed    &     1       & 0.9557 \\
$c2.07650$ & passed    &     1       & 0.5034 \\
$c2.10250$ & passed    &     1       & 0.4120 \\
$c2.11457$ & passed    &     1       & 0.3760 \\
$c2.12435$ & passed    &     1       & 0.4112 \\
$c2.13472$ & passed    &     1       & 0.7506 \\
$c2.15252$ & passed    &     1       & 0.3726 \\
$c2.16837$ & passed    &     1       & 0.0640 \\
$c2.17606$ & passed    &     1       & 0.0954 \\
$c2.18753$ & passed    &     1       & 0.1357 \\
$c3.00580$ & passed    &     1       & 0.6830 \\
$c3.00885$ & passed    &     1       & 0.2725 \\
$c3.01901$ & passed    &     1       & 0.7094 \\
$c3.02968$ & passed    &     1       & 0.3449 \\
$c3.04141$ & passed    &     1       & 0.2200 \\
$c3.05141$ & passed    &     1       & 0.4065 \\
$c3.06631$ & passed    &     1       & 0.9154 \\
\hline 
\end{tabular}
\end{minipage}%
\hfill \hfill
\begin{minipage}{0.45\textwidth}
\centering
\begin{tabular}{l*{10}{c}}
Marker  & stationarity  & start   &   p-value \\
 & test & iteration & \\
\hline  
$c3.08042$ & passed    &     1       & 0.9371 \\
$c3.09748$ & passed    &     1       & 0.1659 \\
$c3.10996$ & passed    &     1       & 0.2741 \\
$c3.11192$ & passed    &     1       & 0.9199 \\
$c3.12647$ & passed    &     1       & 0.5791 \\
$c3.15117$ & passed    &   101       & 0.1036 \\
$c3.16677$ & passed    &     1       & 0.1833 \\
$c3.18180$ & passed    &     1       & 0.4971 \\
$c3.20729$ & passed    &     1       & 0.1290 \\
$c3.22147$ & passed    &     1       & 0.3444 \\
$c4.00012$ & passed    &     1       & 0.6771 \\
$c4.00641$ & passed    &     1       & 0.1561 \\
$c4.02133$ & passed    &   151       & 0.2904 \\
$c4.03833$ & passed    &   151       & 0.0606 \\
$c4.04877$ & passed    &   151       & 0.1526 \\
$c4.05850$ & passed    &     1       & 0.1462 \\
$c4.06923$ & passed    &     1       & 0.1180 \\
$c4.07549$ & passed    &     1       & 0.6850 \\
$c4.07740$ & passed    &     1       & 0.8112 \\
$c4.08930$ & passed    &     1       & 0.5886 \\
$c4.10609$ & passed    &     1       & 0.8891 \\
$c4.11878$ & passed    &     1       & 0.8728 \\
$c4.13171$ & passed    &     1       & 0.9864 \\
$c4.14819$ & passed    &     1       & 0.9607 \\
$c4.15765$ & passed    &     1       & 0.5666 \\
$c4.17684$ & passed    &     1       & 0.7981 \\
$c5.00576$ & passed    &     1       & 0.2602 \\
$c5.01587$ & passed    &     1       & 0.3229 \\
$c5.02900$ & passed    &     1       & 0.5201 \\
$c5.04011$ & passed    &     1       & 0.4648 \\
$c5.05319$ & passed    &     1       & 0.3446 \\
$c5.06820$ & passed    &     1       & 0.3945 \\
$c5.07442$ & passed    &     1       & 0.1545 \\
$c5.08563$ & passed    &     1       & 0.2376 \\
$c5.10428$ & passed    &     1       & 0.9502 \\
$c5.13614$ & passed    &     1       & 0.4708 \\
$c5.14766$ & passed    &     1       & 0.5127 \\
$c5.17570$ & passed    &     1       & 0.0560 \\
$c5.19316$ & passed    &     1       & 0.2139 \\
$c5.20318$ & passed    &     1       & 0.4989 \\
$c5.21319$ & passed    &     1       & 0.8631 \\
$c5.22415$ & passed    &     1       & 0.4174 \\
$c5.23116$ & passed    &     1       & 0.6876 \\
$c5.24997$ & passed    &     1       & 0.7167 \\
$c5.26671$ & passed    &     1       & 0.3110 \\
\hline
\end{tabular}
\end{minipage}}
\end{table}

\begin{figure}[ht]
\centering
\includegraphics[width=0.95\textwidth]{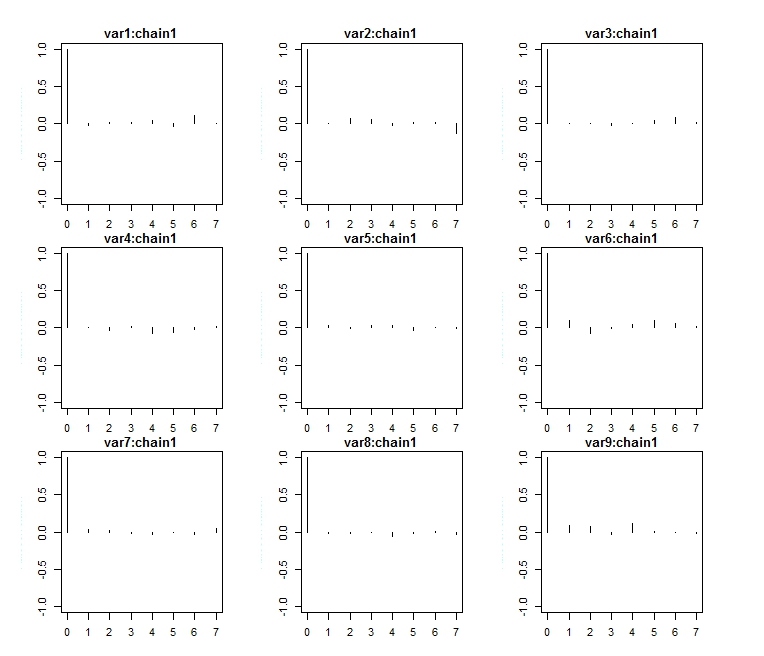}
\caption{The derived samples from the Gibbs sampler within the EM copula are almost independent.\label{autoregressive}}
\end{figure}

Furthermore, we address the sufficient number of samples $Z_\star^{(i)1}, \ldots, Z_\star^{(i)N} $ that is needed to calculate the mean of the expectation. To address this issue, in our simulations and the real data implementation we study the autocorrelation of samples within each variable. In Figure \ref{autoregressive} we show the results of the autoregressive plot for the first $9$ variables in the A.thaliana genotype data. This Figure shows that the obtained samples for each variable are almost independent. Thus, we need few samples to calculate the mean of the expectation. 

\bibliographystyle{Chicago}
\bibliography{ref}

\end{document}